\documentclass[twocolumn,showpacs,secnumarabic,amssymb,nobibnotes,aps,prl]{revtex4-1}
\usepackage{amsmath}
\usepackage{latexsym}
\usepackage{float}
 \usepackage{amssymb}
\usepackage{graphicx}
\usepackage{textcomp}
\usepackage{hyperref}
\usepackage{color,colortbl}

 1
\def\ba{\begin{eqnarray}}
\def\ea{\end{eqnarray}}
\def\be{\begin{equation}}
\def\ee{\end{equation}}
\def\bm{\begin{math}}
\def\me{\end{math}}

\newcommand{\dummy}

\begin{document}

\title{Fractality in Persistence 
Decay and Domain Growth during Ferromagnetic Ordering: Dependence 
upon initial correlation}
\author{ Saikat Chakraborty and Subir K. Das$^{*}$}
\affiliation{Theoretical Sciences Unit, Jawaharlal Nehru Centre for Advanced Scientific Research,
 Jakkur P.O, Bangalore 560064, India}

\date{\today}

\begin{abstract}
\par
Dynamics of ordering in Ising model, following quench to zero temperature, 
have been studied via Glauber 
spin-flip Monte Carlo simulations in space dimensions $d=2$ and $3$. One of the primary 
objectives has been to understand phenomena associated with the persistent spins, viz., 
time decay in the number of unaffected spins, growth of the corresponding pattern and its 
fractal dimensionality, for varying correlation length in the 
initial configurations, prepared at different temperatures, at and 
above the critical value. It is observed that the fractal dimensionality 
and the exponent describing the power-law decay of persistence probability
are strongly dependent upon the relative values of 
nonequilibrium domain size and the initial 
equilibrium correlation length. Via appropriate scaling 
analyses, these quantities have been estimated for quenches from infinite 
and critical temperatures. The above mentioned dependence 
is observed to be less pronounced in higher dimension. In addition to these findings 
for the local persistence, we present results for the global persistence as well.
Further, important observations on the standard domain growth 
problem are reported. For the latter, a controversy in $d=3$, related to the value of the exponent 
for the power-law growth of the average domain size with time, has been resolved.
 
\end{abstract}

\maketitle
\section{\textrm{I} Introduction}
Kinetics of phase transitions \cite{onuki,bray_phase,jones_smsm,bray_per}
remains an active area of research for several decades.
In this area, typically one is interested in the nonequilibrium dynamics related to the evolution 
of a system to a new equilibrium state, having been quenched from a configuration 
prepared outside the coexistence curve to inside it, via the variation of temperature 
($T$), pressure, etc. In this work, our focus is on the paramagnetic to ferromagnetic 
transition \cite{goldenfeld}. When a system is quenched, 
via variation of $T$, from the paramagnetic 
phase to ferromagnetic one, domains rich in 
like spins form and grow with time \cite{bray_phase}. 
Aspects that drew attention of researchers, in this problem, are understanding of 
domain patterns \cite{bray_phase}, growth of domains \cite{bray_phase}, aging 
properties of the evolution \cite{fisher_huse,cor_lippi,jia_suman},
as well as the pattern (and corresponding dynamics) exhibited by atomic magnets (or spins) 
that did not change orientation till time t, referred to as persistent spins \cite{bray_per,satya_sire,%
satya_bray,derrida_hakim,derrida,stauffer,cueille_jpa,cueille_epjb,manoj_1,manoj_2,manoj_3,jain,saharay,jkb,%
paul,blanchard,saikat,gambassi}.
This work deals with issues related to domain growth and persistence.
\par
During the process of ferromagnetic ordering (where the order parameter is 
a nonconserved quantity), the average domain size, $\ell$,
increases as \cite{bray_phase}
\begin{equation}\label{eq1}
\ell \sim t^{\alpha},
\end{equation}
where $\alpha$, the growth exponent, may have dependence upon system dimensionality ($d$) 
based on the order-parameter symmetry. This growth occurs via 
motion and annihilation of defects, facilitated by 
change in orientation of the 
spins, $S_i$, the subscript $i$ being an index related to an atom or spin, 
typically considered to be located on a regular lattice. 
In this work we study the spin-$1/2$ Ising model, to be defined later,
for which defects are the domain boundaries. In this case, 
$S_i$ is a scalar quantity which gets affected only via (complete) flipping or change in 
sign. For this model, the theoretical expectation for $\alpha$
is same in both $d=2$ and $3$.
\par
The persistence probability, $P$, defined as 
the fraction of unaffected spins, typically decays as \cite{bray_per}
\begin{equation}\label{eq2}
P \sim t^{-\theta},
\end{equation}
where $\theta$ is expected to have dependence upon $d$. 
The persistent spins exhibit interesting fractal pattern with 
dimensionality \cite{manoj_2} $d_f$ whose dependence upon $\theta$ will be introduced later.
Unless mentioned otherwise, all our results on this issue
correspond to local persistence, probability
for which, as already mentioned, is calculated by counting 
unaffected ``microscopic'' spins. There has also been interest
in the calculation of such probability by dividing the system 
into blocks of linear dimension $\ell_b$
and counting the persistence of coarse-grained or block spin variables \cite{cueille_jpa,cueille_epjb}.
In the limit $\ell_b \rightarrow a$, the microscopic lattice constant, such block
persistence probability, $P_b$, will correspond to $P$, the ``local or site persistence'' probability.
On the other hand, for $\ell_b \rightarrow \infty$, one obtains ``global persistence'' probability,
further discussion and results for which will be presented later.

\begin{figure}[htb]
\centering
\includegraphics*[width=0.40\textwidth]{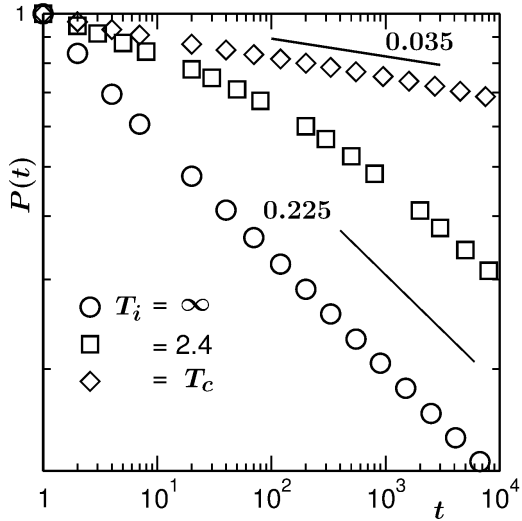}
\caption{\label{fig1} Log-log plots of local persistence probability, $P(t)$, vs $t$, 
for quenches (of the Glauber Ising model) from different values of initial 
temperature $T_i$ ( $\geq T_c$, the critical temperature), to the final value $T_f=0$. 
All results correspond to space dimension $d=2$ and square lattice, with linear 
dimension of the square box being $L=2048$, in units of the lattice constant $a$. 
The lines represent various power-law decays, values of the exponents being mentioned 
in appropriate places.
}
\end{figure}

\par
For Ising model, values \cite{bray_phase,manoj_2,jain}
of $\alpha$, $\theta$ and $d_f$ are accurately estimated
via Monte Carlo (MC) simulations,
in $d=2$, for quenches from initial temperature $T_i=\infty$
to the final value $T_f=0$.
It is reasonably well established \cite{bray_phase,bray_per,%
derrida_hakim,derrida,stauffer,cueille_jpa,cueille_epjb,manoj_1,manoj_2,manoj_3,jain,saikat} 
that, in this case,
the values of $\alpha$, $\theta$ and $d_f$ are $1/2$, 
$0.225$, and $1.58$. However, the conclusions, if exists, 
on the corresponding numbers for $d=3$ are questionable \cite{corberi,amar_family,holzer}. 
Recent focus, on the other hand, for 
persistence as well as for other aspects of coarsening,  
has been on \cite{blanchard,saikat,gambassi,pandit,humayun,sicilia}
quenches from temperatures providing large equilibrium correlation 
length $\xi$. In this context, in a recent work \cite{saikat}, we have explored 
the initial correlation dependence of $\alpha$ and $\theta$. Our observation was,  
while $\alpha$ is insensitive to the variation of $T_i$ (at least in $d=2$), 
$P$ (and thus $\theta$) is strongly influenced 
by the choice of the latter, viz., we obtained for $d=2$ and $3$, 
$\theta=\theta_c \simeq 0.035$ and $\simeq 0.105$ for $T_i=T_c$, the critical 
temperature (see Fig. \ref{fig1} for $d=2$). 
The numbers quoted above are significantly different from those for $T_i=\infty$.
For intermediate temperatures, as seen in Fig. \ref{fig1}, 
two step decays can be noticed.
A slower decay was observed for $\ell < \xi$. The corresponding exponent $\theta_I$ approaches
$\theta_c$ as $\xi \rightarrow \infty$, i.e, when $T_i \rightarrow T_c$. For $\ell \gg \xi$, 
behavior consistent with $T_i=\infty$ was obtained. This implies, 
dynamics of the spins is strongly influenced by the relative 
values of nonequilibrium domain length $\ell$ and 
the equilibrium correlation length $\xi$ in the initial configuration.
The overall time decay of $P$, for all $T_i$, was empirically constructed to be \cite{saikat} 
\begin{equation}\label{eq3}
P(t)x^{2\theta}=A\bigg(\frac{x}{g(x)+x} \bigg)^{\phi};~x=\ell/\xi, 
\end{equation}
with
\begin{equation}\label{eq4}
g(x)=\frac{C_0}{1+C_{1}x^{\psi}},
\end{equation}
where $A$ is the amplitude of the long time decay, $\phi=(\theta-\theta_I)/\alpha$, 
$\psi \simeq 2$, whereas $C_0$ and $C_1$ are dimension dependent constants.
\par
An extension of a study \cite{manoj_2} (via a different model in $d=1$) predicts
\begin{equation}\label{eq5}
d_{f}=d-z\theta ,
\end{equation}
where $z$, to be more formally defined later,
is a dynamical exponent related to the growth of the persistence pattern.
From previous studies \cite{saikat,sicilia}, even though it has been reported 
that the decay of $P$ is disconnected with the growth of $\ell$, 
$z$ and $\alpha$ may be related. Nevertheless, since such a connection is unclear, 
to gain knowledge about the
variation of $d_f$, as a function of $T_i$, estimation of $z$ is needed. 
Even if such a connection exists, as mentioned, the value of $\alpha$
in $d=3$ is not unambiguous.
In this dimension, the  theoretically \cite{bray_phase} expected
value of $\alpha$ $(=1/2)$ disagrees with some computer simulations \cite{cueille_jpa} 
which report numbers close to $1/3$. This difference can 
possibly \cite{corberi} be due to long transient period.
Thus, lengthy simulation runs with large systems are needed. It will be interesting to
see if such long simulation, luck favoring, can provide the theoretically expected value. 
If yes, in that time regime, do we see change in other quantities as well?
\par
In this work, our objective thus, is to estimate $d_f$, $z$, 
$\alpha$ and $\theta$, for $T_i=\infty$ and $T_i=T_c$, in 
space dimensions $2$ and $3$, for quenches to $T_f=0$.
For the ease of reading, in TABLE 1 we provide 
a list of values of these quantities, obtained from computer simulations.
While the ones with asterisks, to the best of our knowledge, will be calculated
(or the simulation results will be shown to be consistent with those theoretical expectations)
for the first time, the numbers appearing inside the parentheses are improvements
over the existing ones that appear outside.
We will start presenting results
with the objective of calculating $d_f$. Other quantities will be needed for this 
purpose and will be estimated in due course.


\begin{table} [ht]
\caption{List of some nonequilibrium exponents for Ising model.}
\centering
\begin{tabular}{ |c|c|c|c|c| }
\hline
Case & $\alpha$ & $z$ & $\theta$ & $d_f$  \\
\hline
$d=2$, $T_i=\infty$ & $1/2$ & $2$ & $0.225$ & $1.58$ $(1.53)$ \\
$d=2$, $T_i=T_c$ & $1/2$ & $2^*$ & $0.035$ & $1.92^*$ \\
$d=3$, $T_i=\infty$ & $1/3$ $(1/2)$ & $2^*$ & $0.18$ $(0.15)$ & $2.65^*$  \\
$d=3$, $T_i=T_c$ & $1/2$ & $2^*$ & $0.105$ & $2.77^*$\\
\hline
\end{tabular}
\end{table}

\begin{figure}[htb]
\centering
\includegraphics*[width=0.45\textwidth]{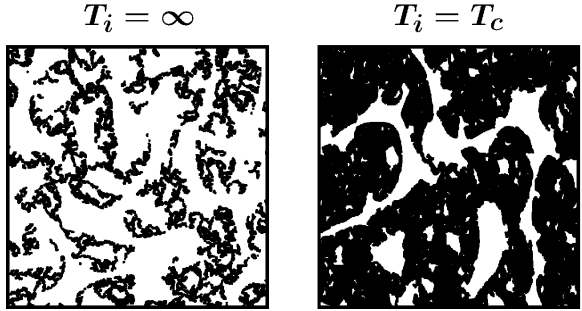}
\caption{\label{fig2} Snapshots of the persistent spins are shown for quenches from 
$T_i=\infty$ and $T_c$, to $T_f=0$. The results correspond to $d=2$, $L=2048$ and 
$t=10^4$ MCS. In both the cases only parts of the boxes are shown where the 
persistent spins are marked in black.
}
\end{figure}

\begin{figure}[htb]
\centering
\includegraphics*[width=0.45\textwidth]{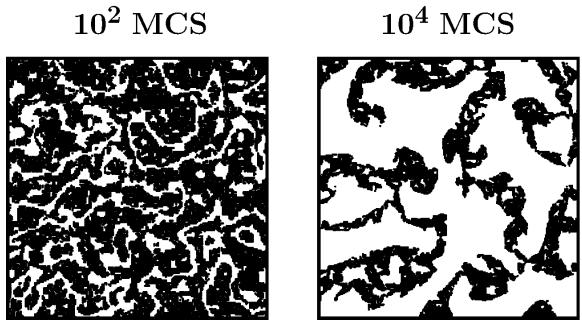}
\caption{\label{fig3} Snapshots of the persistent spins from different times, 
mentioned on the figure, are shown for $T_i=2.4$ and $T_f=0$. Other details 
are same as Fig. \ref{fig2}.
}
\end{figure}

\par
In Fig. \ref{fig2} we show persistence snapshots for $T_i=\infty$ and $T_c$, both from 
$t=10^4$ Monte Carlo steps (MCS), this time unit to be defined soon, for $d=2$ Ising model. It is 
clear that the patterns are different and so, different values of $d_f$ are expected. 
In Fig. \ref{fig3}, snapshots from an intermediate temperature $T_i=2.4$ 
($>T_c$), for $d=2$, are presented. The first frame corresponds to a time falling in 
the slower decay regime of Fig. \ref{fig1} (for the corresponding temperature), 
whereas the second one is from the 
faster decay regime, implying $\ell \gg \xi$. The earlier time snapshot resembles 
the $T_i=T_c$ picture of Fig. \ref{fig2} and the second one has similarity with 
$T_i=\infty$ pattern. This justifies our focus only on these 
two limiting initial temperatures with $\xi=0$ and $\infty$, rather than exploring 
a wide temperature range, to accurately quantify $d_f$ and $z$. 
\par
The rest of the paper is organized as follows. In the next section we describe the 
model and method. Section \textrm{III} provides a brief overview of an earlier work. 
Results are presented in section \textrm{IV}. Finally, section \textrm{V} concludes 
the paper with a brief summary and outlook.

\section{\textrm{II} Model and Method}
\par
As already mentioned, we study the Ising model \cite{goldenfeld},
on square or simple cubic lattice systems, depending upon the dimensionality, with nearest neighbor 
interactions. The Hamiltonian for the model is given by 
\begin{equation}\label{eq6}
H=-J\sum_{<ij>} S_{i}S_{j};~S_i=\pm1,
\end{equation}
where $J$ is the interaction strength ($>0$) and $<ij>$ implies interaction 
among nearest neighbors. The values of $T_c$ for this model 
in $d=2$ and $3$ are respectively \cite{d_landau} 
$\simeq2.27 J/k_{B}$ and $\simeq 4.51 J/k_{B}$, 
$k_B$ being the Boltzmann constant. 
\par 
Kinetics in this model was introduced via Glauber 
spin-flip mechanism \cite{d_landau,glauber}. In this 
MC approach, a trial move consists of changing the sign of a randomly chosen spin. 
Since our quenches were done to $T_f=0$, a move was accepted only if 
it had reduced the energy. Needless to say, initial configurations were prepared 
at nonzero $T$ values. In that case, the Metropolis criterion for the acceptance of 
a move was implemented via appropriate calculation 
of the Boltzmann factor \cite{d_landau} and 
its comparison with a random number, ranging between $0$ and $1$, whenever the move 
brought an increment in the energy. 
For preparation of initial configurations at 
temperatures very close to $T_c$, in addition to the Glauber mechanism, 
we have applied Wolff algorithm \cite{wolff} as well, which facilitates faster equilibration.
Time, in our simulations are measured in units of MCS, each MCS consisting of $L^d$ 
steps, $L$ being the linear dimension of a square or cubic box. Periodic boundary 
conditions were applied in all directions. 
Final results are presented after averaging over multiple initial realizations, 
the number ranging from 20 to 70. In $d=2$ all results are for $L=2048$.
In $d=3$, the results for $T_i=\infty$ are for $L=512$ and for 
$T_i=T_c$, we presented results from $L=400$ and $256$.
\section{\textrm{III} An Overview of the Background On fractality of persistence pattern}
\par
In this section we provide a discussion on the theoretical background
for fractality of the structures formed by persistent spins, following 
the work by Manoj and Ray \cite{manoj_2}.
\par 
 From a density correlation function, $D(r,t)$, isotropic in an unbiased system,
total mass or number of particles in a circular or spherical 
(depending upon dimensionality) region of radius $R$ can be obtained as
\begin{equation}\label{eq7}
M(R,t) \sim \int_0^R D(r,t)r^{d-1}dr,
\end{equation}
$r$ (= $|{\vec{r}}|$) being the scalar distance of a point in that region from the central one. 
An appropriate correlation function in the present context is
\begin{equation}\label{eq8}
D(r,t)=\frac{\langle \rho(\vec{r_0},t)\rho(\vec{r_0}+\vec{r},t)\rangle}{\langle {\rho(\vec{r_0},t)} \rangle},
\end{equation}
with $\rho$ being unity at a space point if the spin there did not flip till 
time $t$ and zero otherwise. The average order 
parameter for the persistent pattern is
\begin{equation}\label{eq9}
\langle \rho(\vec{r},t)\rangle=\frac{\int {d\vec{r}\rho(\vec{r},t)}}{\int{d\vec{r}}}=P(t).
\end{equation}
This being a nonconserved (time dependent) quantity and, since, in the definition of $D(r,t)$, the 
average value is not subtracted from $\rho$, decorrelation here means, decay of $D(r,t)$ 
to a ``non-zero'' value (=$P(t)$), for $t<\infty$. The distance, $\ell_{p}(t)$, at which $D(r,t)$
reaches this plateau is the characteristic length scale of the pattern. In that 
case, there may exist scaling of the form 
\begin{equation}\label{eq10}
\frac{D(r,t)}{P(t)} \equiv f(r/l_p).
\end{equation}
For $x$ ($\equiv r/l_p$) $> 1$, $f$ should be unity. On the other hand, for fractal dimension 
$d_f$ and $x<1$, one should have
\begin{equation}\label{eq11}
f(x) \sim x^{d_f-d},
\end{equation}
since
\begin{equation}\label{eq12}
M \sim x^{d_f}.
\end{equation}
\par
Considering that $P(t)$, the plateau value, decays in a power-law fashion, a power-law behavior
of $f(x)$ is indeed expected, once scaling is achieved. A continuity, 
at $r=\ell_p$, in such a situation demands 
\begin{equation}\label{eq13}
t^{(d_f-d)/z}=t^{-\theta},
\end{equation}
providing Eq. (\ref{eq5}), where $z$ is the dynamic exponent characterizing the growth 
of the persistence pattern, mentioned before, as
\begin{equation}\label{eq14}
\ell_p \sim t^{1/z}.
\end{equation}
For this model, as mentioned, value of $\alpha$ has been estimated \cite{saikat} for various $T_i$
values in $d=2$. However, a priori it is unclear whether there is a general validity of the relation
\begin{equation}\label{eq15}
z\alpha=1.
\end{equation}
Then it is necessary to calculate both $z$ and $\theta$, for correlated and 
uncorrelated initial configurations, to validate Eq. (\ref{eq5}). 
On the other hand, as already mentioned, the value 
of $\alpha$ is ambiguous in $d=3$.

\section{\textrm{IV} Results}
\begin{figure}[htb]
\centering
\includegraphics*[width=0.4\textwidth]{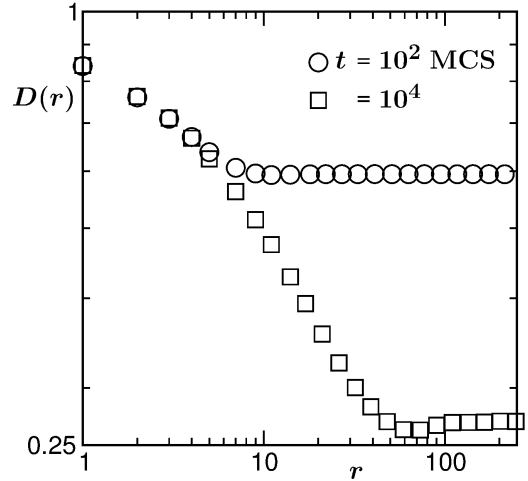}
\caption{\label{fig4} Density correlation functions, $D(r,t)$, related to the persistent 
spins, are plotted vs $r$. Results are presented from two different times, for $T_i=2.4$ 
and $T_f=0$. The system dimensionality is $d=2$ and value of $L$ is $2048$.
}
\end{figure}

\par
In Fig. \ref{fig4} we show $D(r,t)$ as a function of $r$ for $T_i=2.4$, from
two different times, mentioned on the figure, for $d=2$. As expected, the correlation 
function decays to different constant value, $P(t)$, at different length $\ell_p$, for different times.
Before decaying to the plateau, the early time data appear to obey a power-law. The later 
time data, for smaller $r$, follows the same power-law before crossing over to 
another, faster, power-law decay. This implies, there exist two length scales in 
the problem, below and beyond the equilibrium scale $\xi$. 
Inside the larger structure, the small length scale structure
remains hidden, which will become irrelevant in the long time limit.
For $\xi=\infty$, i.e., $T_i=T_c$, however, the latter will be the only structure and remain for ever. 
The exponent for large $r$, for $T_c<T_i<\infty$ and 
$t \gg 0$, should be related to the $d_f$ value for 
$T_i=\infty$ case whereas, in case of small $r$, the exponent should be 
connected to $d_f$ for $T_i=T_c$ case. Below we focus on these two cases, i.e., $T_i=\infty$ 
and $T_i=T_c$, separately, first for $d=2$, followed by $d=3$.
As need occurs, we will present results related to 
$\alpha$, $\theta$ and $z$.

\begin{figure}[htb]
\centering
\includegraphics*[width=0.45\textwidth]{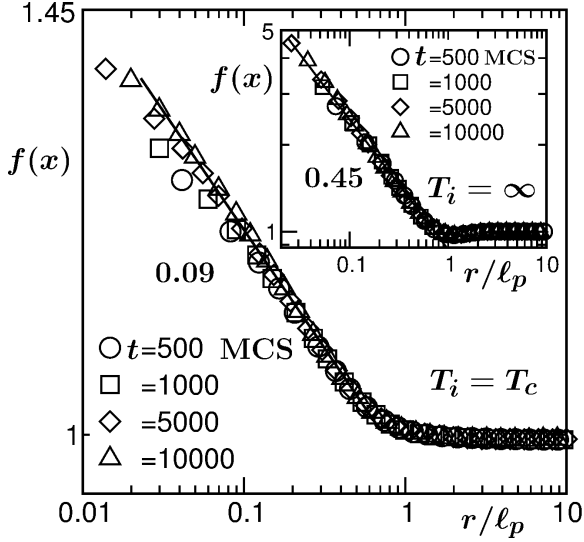}
\caption{\label{fig5} Scaling analysis of $D(r,t)$ for the $d=2$ Ising model,
with $T_i=T_c$ and $T_f=0$, where $f(x)$ is plotted vs $x=r/\ell_p$,
using data from different times after the quench, on log-log scale. The solid 
line corresponds to a power-law decay with an exponent $0.09$. 
Inset: Same as the main frame but for $T_i=\infty$. The solid line here has 
the power-law decay exponent $0.45$. The value of $L$ is $2048$ for all the 
results.
}
\end{figure}

\par
In the main frame of Fig. \ref{fig5} we present a 
scaling exercise \cite{manoj_2,jain} for $D(r)$ where 
we have plotted $f(x)$ as a function of $x$, using data from different times after quench, 
for $T_i=T_c$ and $d=2$. Scaling appears good and gets better with the progress 
of time. On this log-log plot, look of the data appear, before 
decaying to unity, linear, implying a power-law decay. The exponent appears 
to be $\simeq 0.09$.
In the inset of this figure, we show analogous exercise 
for $T_i=\infty$. Even though this case in this dimension was studied by Jain 
and Flynn \cite{jain}, for the sake of comparison and completeness, we present it here
from our own simulations. 
In this case, the exponent for the power-law decay appears consistent with $0.45$. 
Then, in $d=2$, for $T_i=\infty$, the fractal dimensionality is $1.55$ 
and for $T_i=T_c$, the number is $1.91$, if Eq. (\ref{eq11}) is valid.

\begin{figure}[htb]
\centering
\includegraphics*[width=0.45\textwidth]{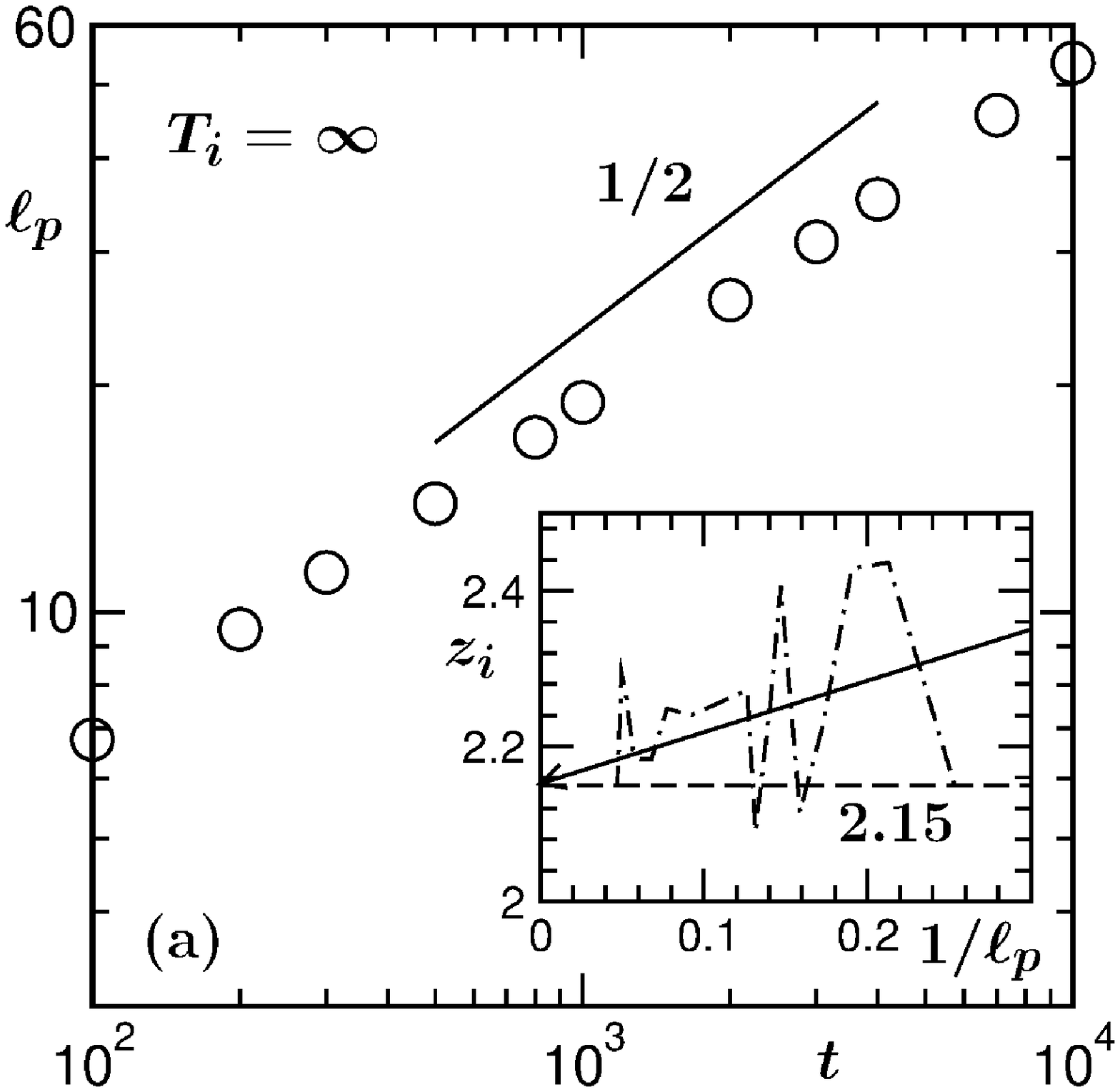}
\vskip 0.4cm
\includegraphics*[width=0.45\textwidth]{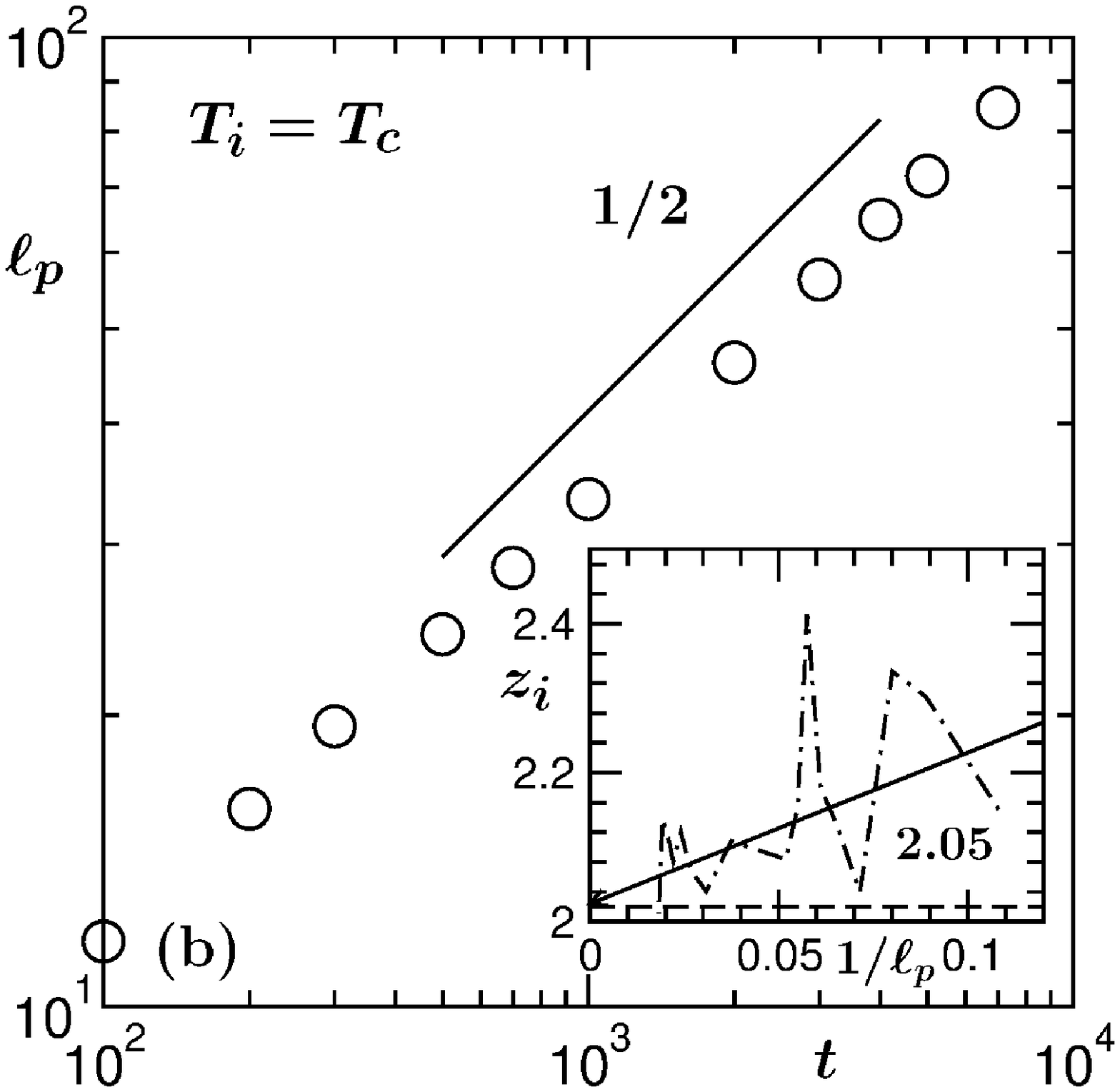}
\caption{\label{fig6}
(a) Log-log plot of persistence length scale, $\ell_p$, 
as a function of $t$, for $d=2$ Ising model, following quench from 
$T_i=\infty$ to $T_f=0$, with $L=2048$.
Inset: Instantaneous exponent, $z_i$, obtained using the data in main 
frame, is plotted vs. $1/\ell_p$.
(b) Same as (a), but here $T_i=T_c$. The solid lines in the main frame of both (a) and 
(b) correspond to power-law growths with exponent $1/2$. In the insets, the horizontal 
dashed lines correspond to our estimates for $z$, whereas solid lines are guides to the eyes.
}
\end{figure}

\par
In Fig. \ref{fig6} we show the plots of $\ell_p$ vs $t$ in $d=2$, for 
(a) $T_i=\infty$ and (b) $T_i=T_c$, on log-log scales. In both 
the cases the data appear consistent with $z=2$, validating Eq. (\ref{eq15}) 
(note that $\alpha$ is established to be $1/2$ in $d=2$).
Nevertheless, we intend to make more accurate quantification. For this purpose, 
in the insets of these figures we have shown instantaneous 
exponents (dash-dotted lines), $z_i$, calculated as \cite{huse} 

\begin{equation}\label{eq16}
 \frac{1}{z_i}=\frac{d\ln {\ell_p}}{d\ln t},
\end{equation}
vs $1/\ell_p$. In both the cases we obtain the value of $z$ via linear 
extrapolation (see the consistency of the simulation data with the solid line) to $\ell_p=\infty$.
For $T_i=\infty$, from this exercise, we quantify $z=2.15$  
and for $T_i=T_c$, we obtain the number $z=2.02$ (see the dashed 
horizontal lines).
These numbers, in addition to verifying Eq. (\ref{eq15}), are also 
consistent with the numbers obtained via least square fitting of the
$\ell_p$ vs $t$ data to the form
\begin{equation}\label{eq17}
 \ell_p=\ell_z^0+{A_z}t^{1/z},
\end{equation}
where $\ell_z^0$ and $A_z$ are positive constants. 
This consistency may imply, early time corrections 
to the exponents are insignificant. Note here that, 
in absence of any correction, one expects \cite{amar,das}
\begin{equation}\label{eq18}
\frac{1}{z_i}=\frac{1}{z}\Bigg[1-\frac{\ell_z^0}{\ell_p} \Bigg],
\end{equation}
a linear behavior of $1/z_i$, when plotted vs $1/\ell_p$, with slope $-\ell_z^0/z$.
A positive slope in both the insets is due to the fact that we have 
presented inverse of the quantity discussed in Eq. (\ref{eq18}).
Using these values of $z$, and 
numbers for $\theta$, mentioned earlier, 
in Eq. (\ref{eq5}), we obtain $d_f=1.93$ for $T_i=T_c$ 
and $d_f=1.51$ for $T_i=\infty$. These values, within computational errors, are consistent 
with the conclusions from Fig. \ref{fig5}. 
Next we present results from $d=3$. 
\par
\begin{figure}[htb]
\centering
\includegraphics*[width=0.45\textwidth]{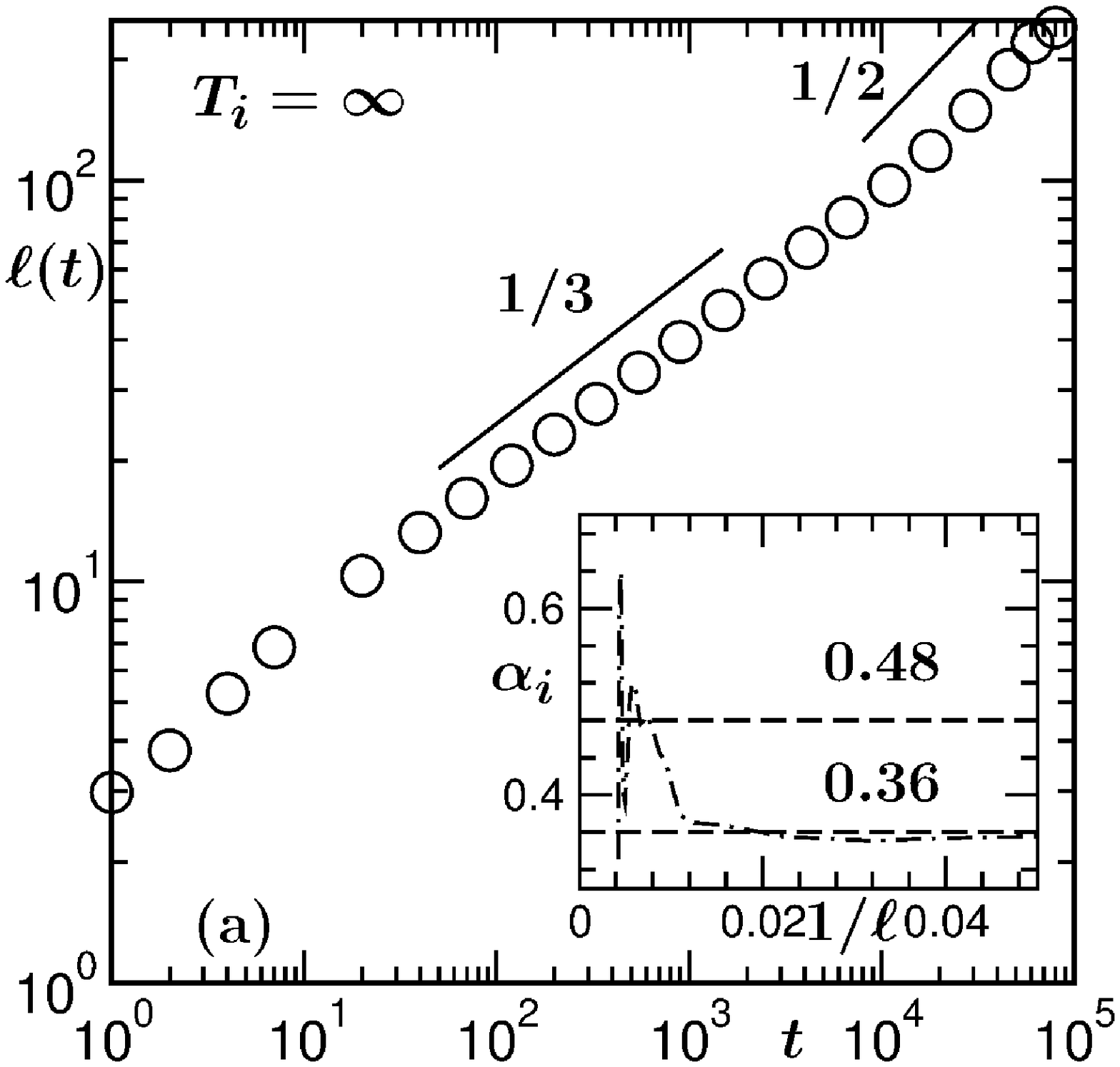}
\vskip 0.4cm
\includegraphics*[width=0.45\textwidth]{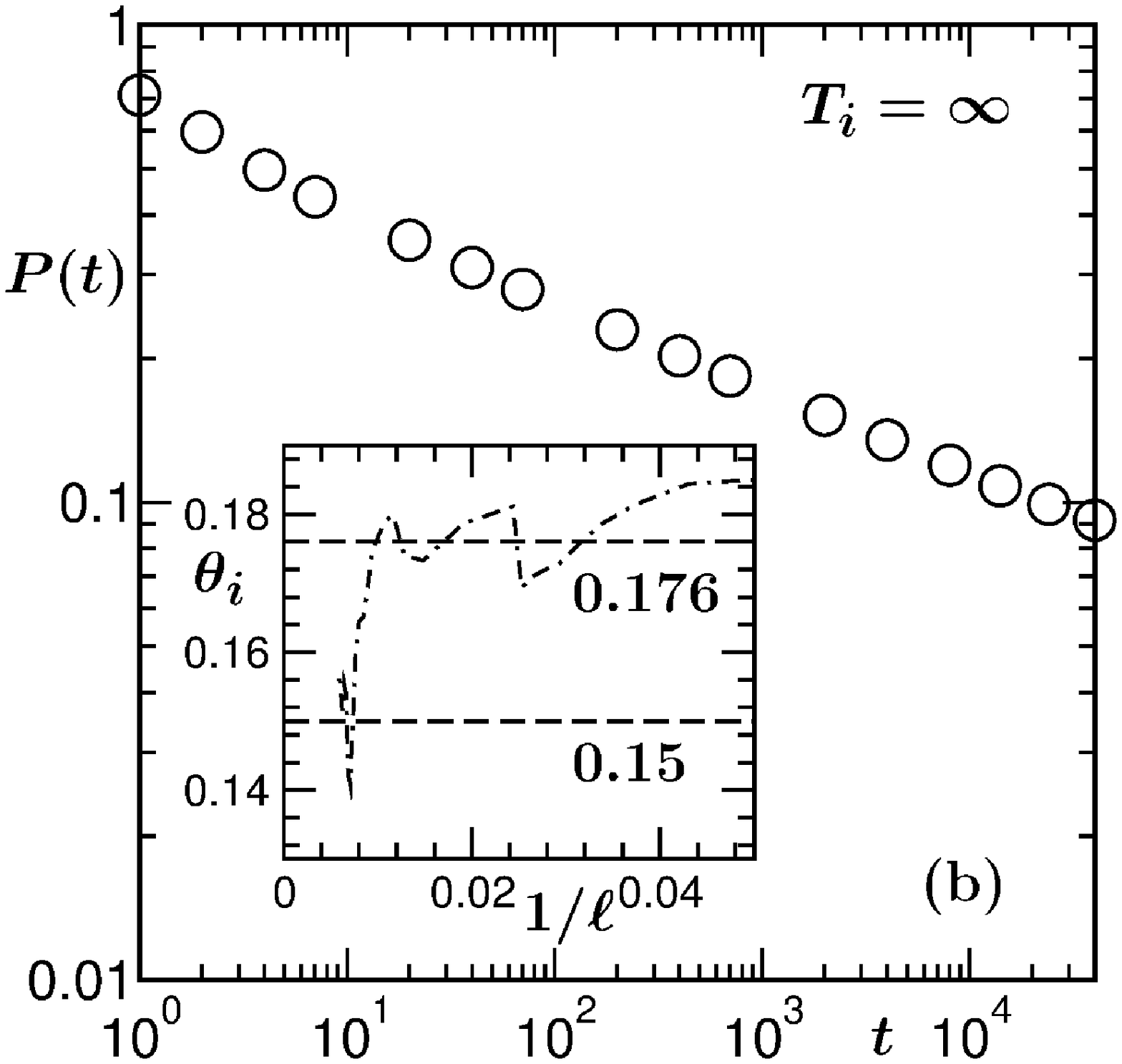}
\caption{\label{fig7}(a) Log-log plot of $\ell$ vs t,
in $d=3$, for $T_i=\infty$. The solid lines correspond to different power
laws, exponents for which are mentioned. The inset shows instantaneous 
exponent $\alpha_i$, as a function of $1/\ell$. The dashed horizontal lines 
represent exponent values $0.36$ and $0.48$. 
(b) Log-log plot of $P(t)$ vs $t$, for $d=3$ and $T_i=\infty$. The inset shows 
instantaneous exponent $\theta_i$ vs $1/\ell$. Horizontal dashed lines 
are for $\theta=0.176$ and $0.15$.
}
\end{figure}

In $d=3$, we start by presenting results for the growth of 
$\ell$, considering the controversy \cite{cueille_jpa,saikat,corberi} on the 
value of $\alpha$ discussed above. In $d=2$, we avoided presenting results on this aspect
with the understanding that the issue there is well settled. 
Nevertheless, in the context of global persistent decay, we will make indirect
conclusion about it. Here note that the estimation of $\ell$ was 
done from the first moment of domain  size distribution, $p(\ell_d,t)$, as
\begin{equation}\label{eq19}
\ell(t)=\int{{\ell_d}p(\ell_d,t)d{\ell_d}},
\end{equation}
where $\ell_d$ is the distance between two domain boundaries in a particular 
direction. The main frame of Fig. \ref{fig7}(a) shows a plot of $\ell$ vs $t$,
on log-log scale, for quenches of the $d=3$ Ising system from $T_i=\infty$ to
$T_f=0$. There exists an intermediate  time regime, extending over more than two 
decades, during which the simulation data show consistency with an exponent 
$\alpha=1/3$, in agreement with previous results \cite{cueille_jpa}. However, 
as discussed and a trend demonstrated in Ref. \cite{corberi},
the discrepancy in the earlier reports from the 
theoretical number $1/2$ can be due to long transient.
Thus, long simulation runs with large systems are necessary. 
We have simulated a system with $L=512$ for time longer 
than any of the previous works, to the best of our knowledge. 
Indeed, it appears that the long time behavior, over the longest time decade 
in the presented time range,
is consistent with $\alpha=1/2$. In the inset of this figure we show the instantaneous 
exponent 
\begin{equation}\label{eq20}
\alpha_i=\frac{d\ln \ell}{d\ln t} , 
\end{equation}
as a function of $1/\ell$. This provides an accurate picture, the long time 
exponent being within 5$\%$ of the theoretical value. One may then ask, is the 
value of $\theta$ going to change, beyond this crossover time? 
Even if $\theta$ has no dependence on the value of $\alpha$,
such a change may still occur.
Note that conclusion on the value of $\theta$, in earlier works \cite{saikat,manoj_1}, were drawn 
from runs shorter than this.
Indeed, a jump in $\theta_i$, calculated from
\begin{equation}\label{eq21}
\theta_i=-\frac{d\ln P}{d\ln t} ,
\end{equation}
occurs (see Fig. \ref{fig7}(b) and the corresponding inset) 
from an early time value of $\simeq 0.18$ to 
$\simeq 0.15$. This may, of course, be 
due to statistical or other reasons. However, since the jump in $\alpha_i$ occurs 
around the same time as the one for $\theta_i$ and fluctuation is seen around stable 
mean values, in both $\alpha_i$ and $\theta_i$, we accept this as the correct 
number for $\theta$ in the asymptotic time limit.
\par
\begin{figure}[htb]
\centering
\includegraphics*[width=0.45\textwidth]{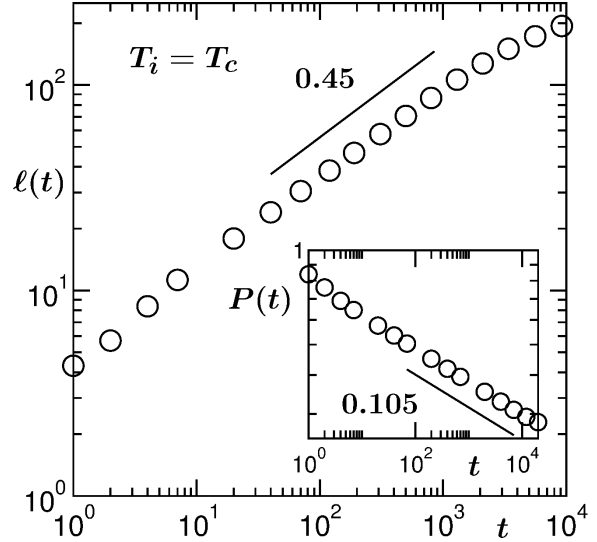}
\caption{\label{fig8} Log-log plot of $\ell$ vs $t$,
for $d=3$, $L=400$ and $T_i=T_c$. The solid line corresponds to a power-law 
growth with exponent $0.45$. The inset shows a plot of $P$ vs $t$, on log-log
scale. The solid line there represents a power-law decay with exponent $0.105$.
}
\end{figure}

Whether due to lattice anisotropy \cite{cueille_jpa} or anything else, 
the solution to overcome such long transient is certainly related to being able to access 
large length scales. For $T_i$ close to $T_c$, since this is 
automatically the case, due to large initial correlation,
we expect an enhanced value of $\alpha$ from early time. Corresponding 
$\ell$ vs $t$ data are presented in Fig. \ref{fig8}. On the log-log scale, this data 
set shows consistency with $\alpha=0.45$. Here we mention that study for $T_i=T_c$
has additional problems related to longer equilibration time at the initial temperature and stronger 
finite-size effects during the nonequilibrium evolution \cite{saikat}. The latter remark can be
appreciated from the plot in Fig. \ref{fig8} where a bending of the data set (from the power-law behavior)
is visible from $t=10^3$. This should be compared with the corresponding data 
in Fig. \ref{fig7}(a) for $T_i=\infty$.
Thus, accessing very large length scales, without finite-size effects,
for $T_i=T_c$, is extremely difficult. The $P$
vs $t$ data, shown in the inset of Fig. \ref{fig8}, exhibit consistency \cite{saikat} with 
$\theta_{c} \simeq 0.105$. Since, $\alpha$ is very close to $1/2$ already, we do not 
expect much change in $\theta_c$ even in true asymptotic length or time limit. The 
calculations of $\alpha_i$ and $\theta_i$ in this case provide numbers consistent 
with the ones quoted above.
\par
Next, we come back to the issue of fractality. For $d=3$
Ising model, our results in this context are entirely new irrespective 
of the value of $T_i$.

\begin{figure}[htb]
\centering
\includegraphics*[width=0.45\textwidth]{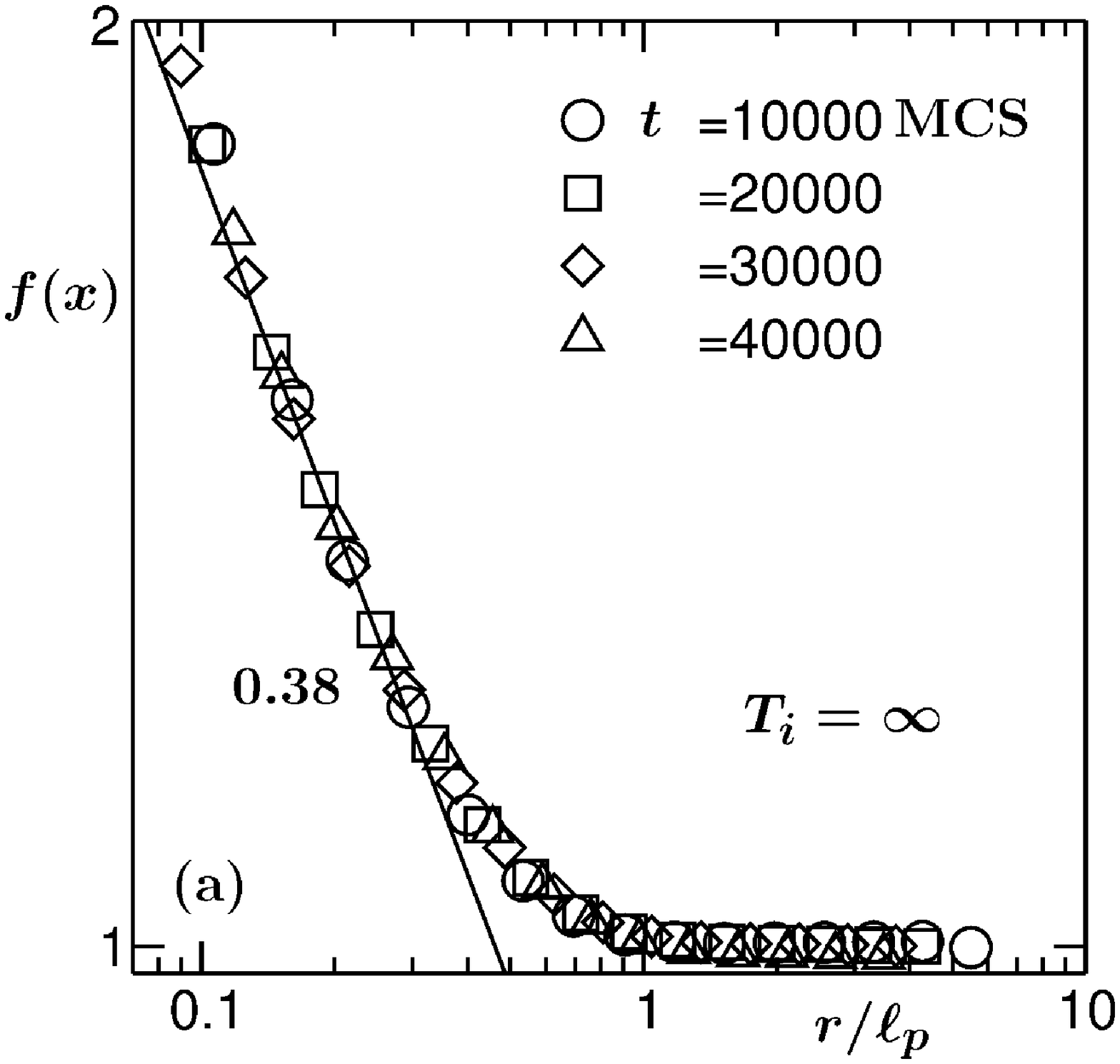}
\vskip 0.4cm
\includegraphics*[width=0.45\textwidth]{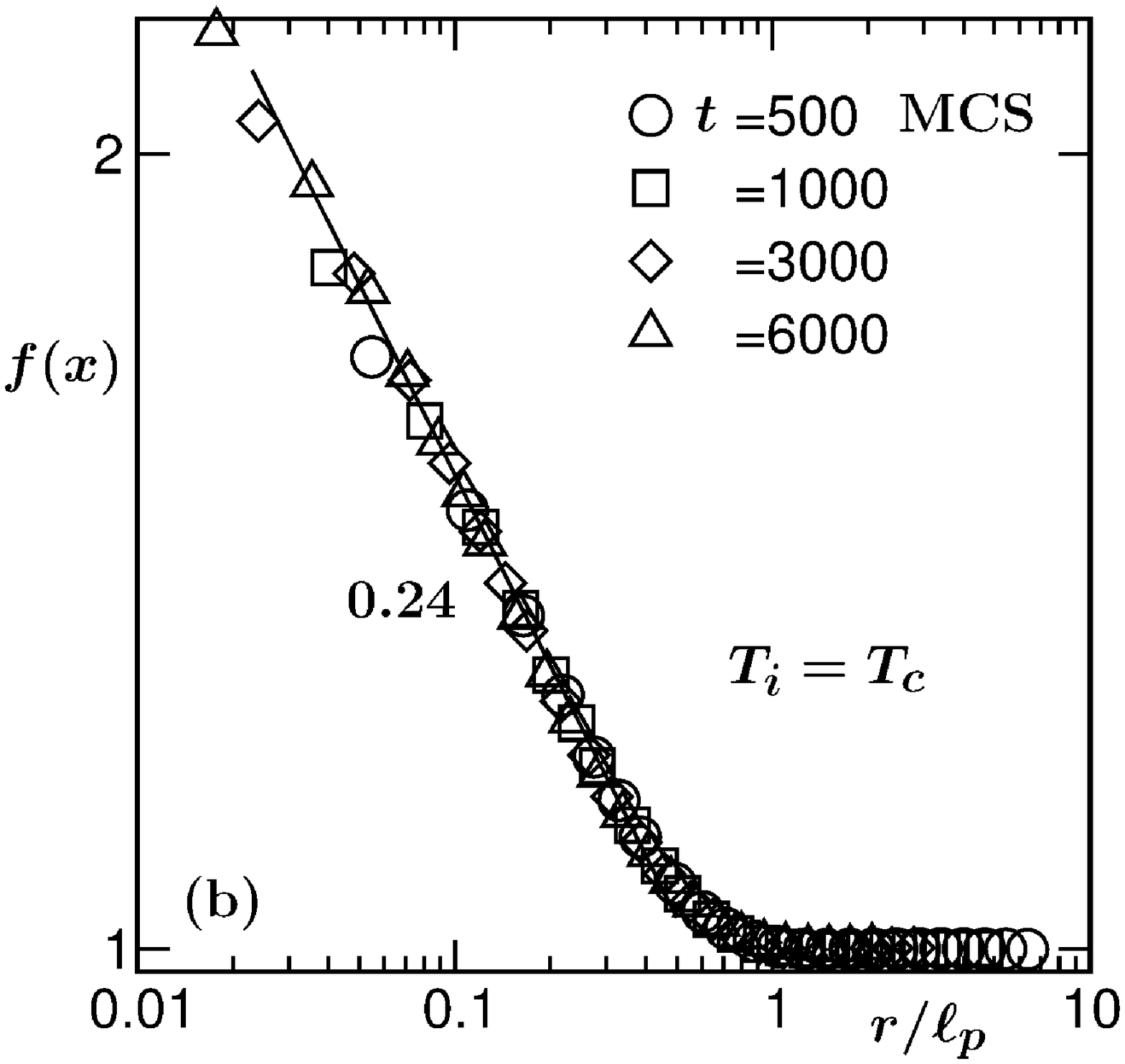}
\caption{\label{fig9}(a) Scaling function $f(x)$ is plotted vs $x$, for $d=3$, 
$T_i=\infty$ and $T_f=0$, using data from few different times. The solid line 
has a power-law decay with exponent $0.38$. 
(b) Same as (a) but, 
instead of $T_i=\infty$, we present data for $T_i=T_c$. Here the 
solid line has power-law decay exponent $0.24$. The results were obtained 
for simple cubic lattice with $L=256$.
}
\end{figure}
\begin{figure}[htb]
\centering
\includegraphics*[width=0.45\textwidth]{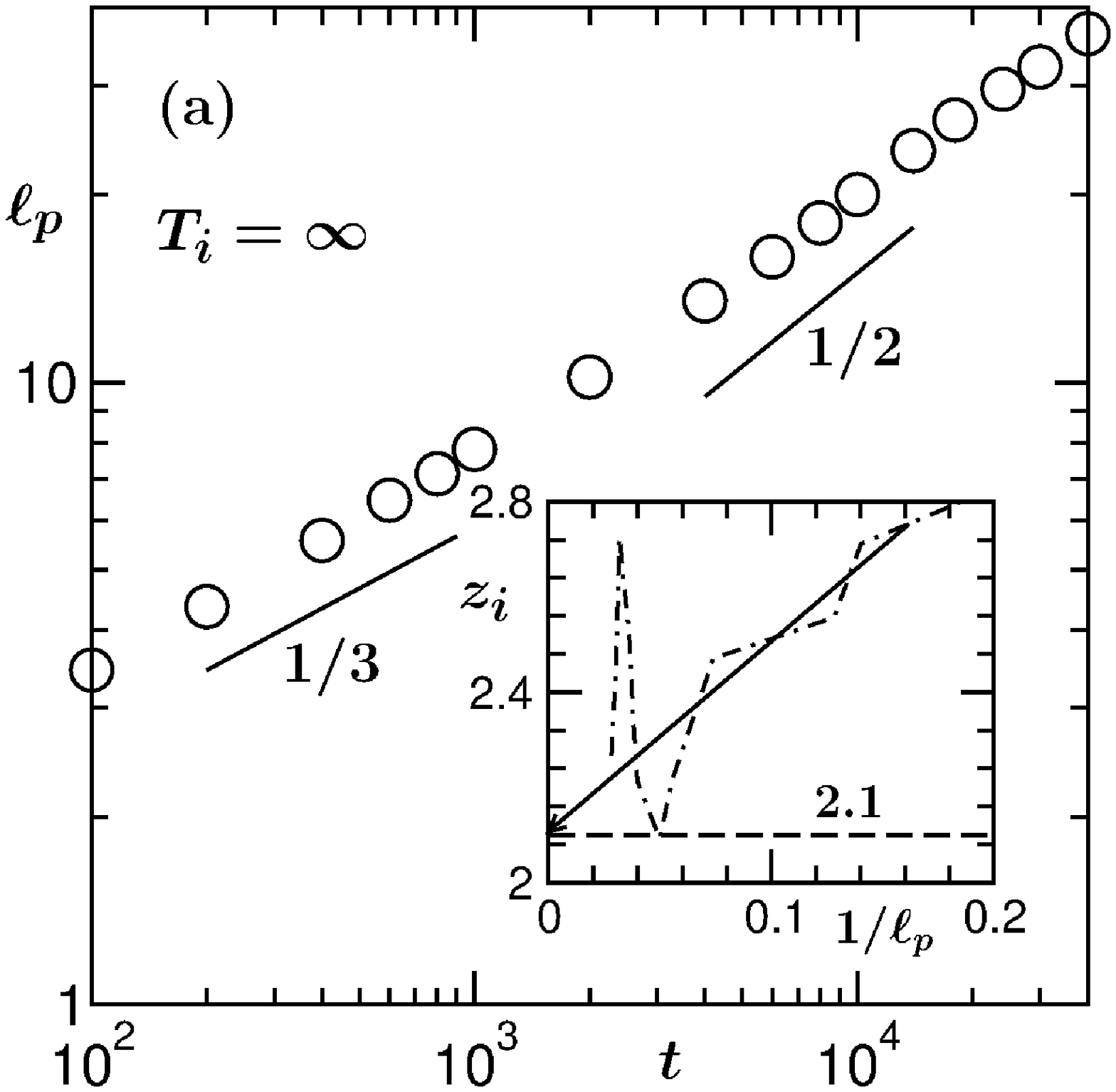}
\vskip 0.4cm
\includegraphics*[width=0.45\textwidth]{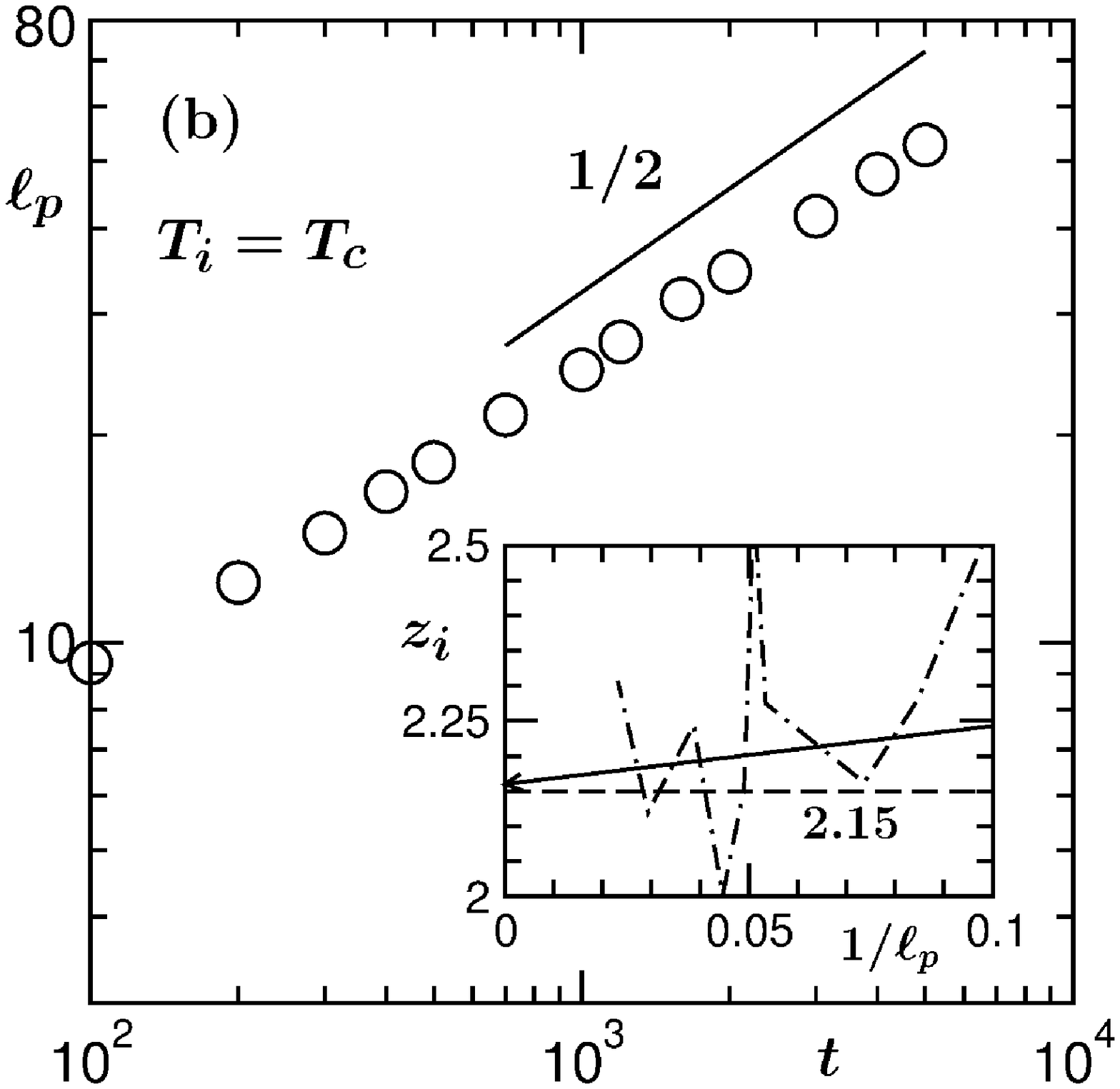}
\caption{\label{fig10} 
(a) Log-log plot of persistence length scale, $\ell_p$, 
as a function of $t$, for $d=3$ Ising model, following quench from 
$T_i=\infty$ to $T_f=0$, with $L=512$. 
Inset: Instantaneous exponent, $z_i$, obtained using the data in main 
frame, is plotted vs. $1/\ell_p$.
(b) Same as (a), but here $T_i=T_c$ and $L=256$. The solid lines in the main frame of both (a) and 
(b) correspond to power-law growths with exponents mentioned there. The horizontal 
dashed lines in the insets correspond to our estimates for $z$ and the solid
lines there are guides to the eyes.
}
\end{figure}

\par
Fig. \ref{fig9}(a) is analogous to the 
inset of Fig. \ref{fig5} ($T_i=\infty$ results for $f(x)$) but for $d=3$. 
The corresponding $f(x)$ vs $x$ scaling plot for $T_i=T_c$ in $d=3$ is presented in 
Fig. \ref{fig9}(b). Again, for 
both $T_i=\infty$ and $T_i=T_c$, good data collapse are obtained for 
results from different times, in these scaling plots. 
For both values of $T_i$, we have used data sets lying in 
the time ranges that provide consistency with the expected theoretical
number for $\alpha$.
In the relevant region, the $T_i=T_c$ results 
have power-law decay with exponent $0.24$. 
In case of $T_i=\infty$, the value of this exponent is approximately $0.38$. 
These numbers imply $d_f=2.76$ and $2.62$ for $T_i=T_c$ and $T_i=\infty$, 
respectively.
\par
In Fig. \ref{fig10} we show $\ell_p$ vs $t$ plots from $d=3$ for (a) $T_i=\infty$ 
and (b) $T_i=T_c$. In the long time 
limit, the results, in both (a) and (b), appear consistent with growth having $z=2$. 
This is in agreement with 
Eq. (\ref{eq15}). From the log-log plot for $T_i=\infty$, like $\ell$ vs $t$,
a long time transient is clearly visible.
To quantify $z$ more accurately (in the $t \rightarrow \infty$ limit), 
for both $T_i=\infty$ 
and $T_i=T_c$, we have shown the instantaneous exponents, vs $1/\ell_p$, in the 
insets. From there, we extract $z=2.1$ for 
$T_i=\infty$ and $2.15$ for 
$T_i=T_c$. Alongwith the above mentioned numbers for $z$, using the values 
of $\theta$ for quenches from $T_i=T_c$ and $T_i=\infty$, we obtain 
$d_f \simeq 2.78$ and $2.68$. These numbers are consistent 
with those obtained from the scaling plots in Fig. \ref{fig9}, 
providing higher confidence on our estimation of $\theta$ from long 
time limit, for $T_i=\infty$. An interesting exercise here would have been 
to plot $z_{i}\alpha_i$ vs $t$. However, a constant value of unity cannot be obtained 
because of the fact that $\ell$ and $\ell_p$ have different initial off-sets.
This is evident from the pictures in the insets of Fig. \ref{fig7}(a) and Fig. \ref{fig10}(a).
While for the time dependence of $\ell$, a long transient with $\alpha \simeq 1/3$ is 
visible, this is not so for the time dependence of $\ell_p$. Thus, because of the off-set related reason 
mentioned above, $ z_{i}\alpha_i=1$ is expected to be valid only in the $t \rightarrow \infty$ limit.
 
\par
Finally, we turn our attention to the block persistence which was 
introduced by Cueille and Sire \cite{cueille_jpa}. The corresponding probability 
$P_b$, as already mentioned, is related to the change in the order-parameter variable obtained by 
coarse-graining the site or microscopic spin variables over a block of linear 
size $\ell_b$. It is expected that the decay of this probability will be significantly 
slower than the site or local persistence probability, to which the 
former should cross over only for $\ell>\ell_b$. 
This two time-scale behavior is desirable by considering 
that, in the early time regime, a slower decay is forced by the fact that a sign
change in block spin variable happens only when $\ell$ becomes comparable to $\ell_b$ and
in the large $\ell$ limit, the blocks effectively appear as sites. It is 
expected then that a scaling should be obtained as \cite{cueille_jpa}
\begin{equation}\label{eq22}
P_{b}{\ell_{b}}^{\theta_0/\alpha} \equiv h(t/{\ell_b}^{1/\alpha}),
\end{equation}
where $\theta_0$ is the exponent of the early part of the decay or global 
persistence exponent in the sense that when $\ell_b \rightarrow \infty$, 
this is the only exponent. In $d=2$, we will see that 
the best scalings, in accordance 
with Eq. (\ref{eq22}), are obtained for 
$\alpha=1/2$, irrespective of 
the value of $T_i$. In $d=3$, on the other hand, due to 
long transient in the dynamics, we avoid presenting these results.
\par
In addition to the above mentioned understandings, calculation of 
persistence probability via such blocking may have advantage for quenches to nonzero 
temperature. 
Note that for $T_f \ne 0$, thermal fluctuation from 
bulk of the domains affects the calculation when done via 
standard method. Considering that domain growth occurs essentially due to spin flips
along the domain boundaries, in the calculation of $P$, dynamics inside the domains needs
to be discarded. In a method, prescribed by Derrida \cite{derrida}, this is done by simulating an
ordered system, alongside the coarsening one, and subtracting 
the common flipped spins, identifiable as the bulk flips,
between the two systems, from the total, 
thus sticking to the effects of only the
boundary motion. In the block spin method, if $\ell_b$ is significantly larger
than $\xi$ at $T_f$, thermal fluctuations will not alter the sign of block spins
and in the large $\ell$ ($>\ell_b$) limit, as previously stated, one 
expects the decay to be consistent with
local persistence. This saves computational time for simulating the
additional systems with ordered configurations.
\begin{figure}[htb]
\centering
\includegraphics*[width=0.45\textwidth]{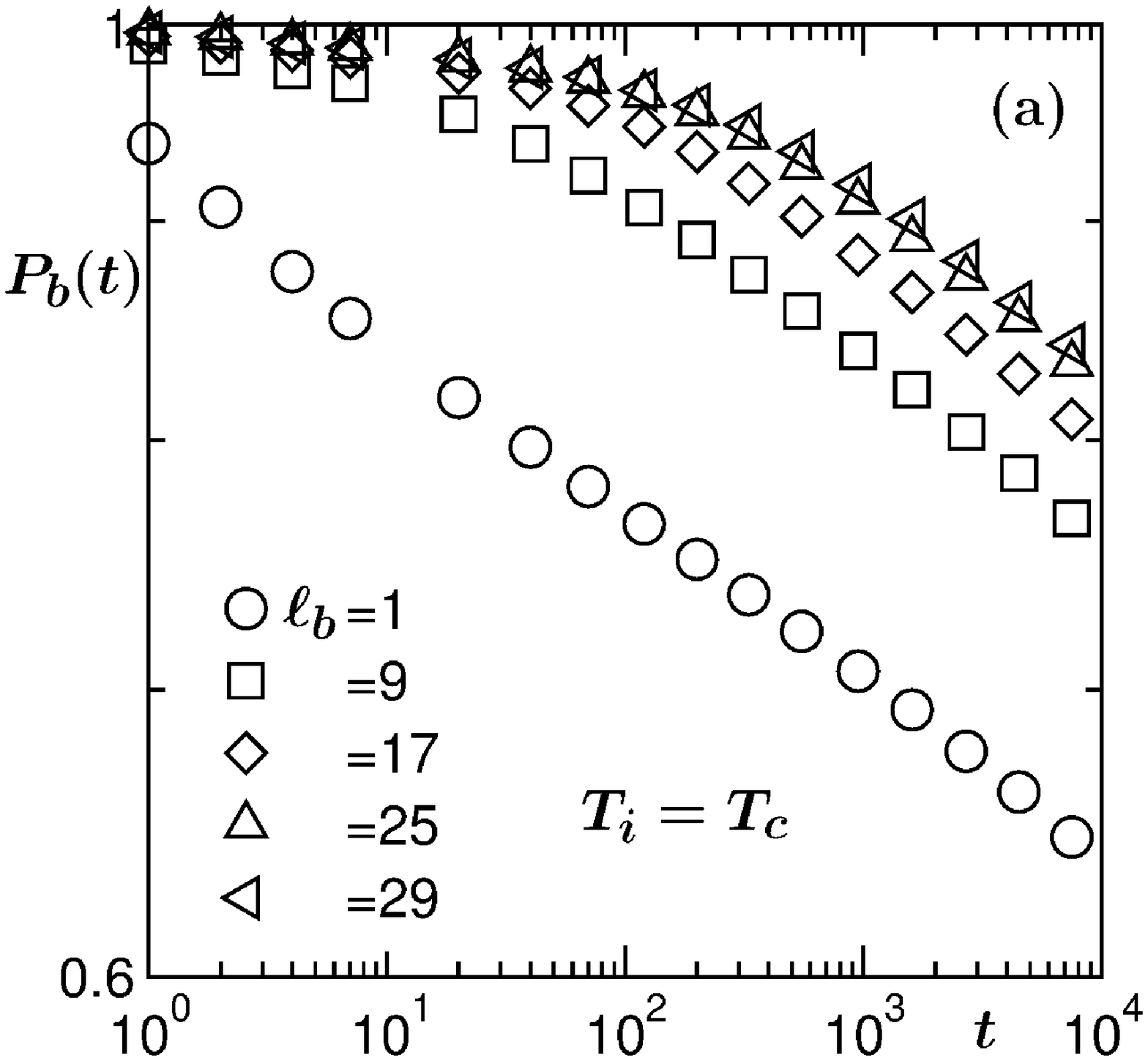}
\vskip 0.4cm
\includegraphics*[width=0.45\textwidth]{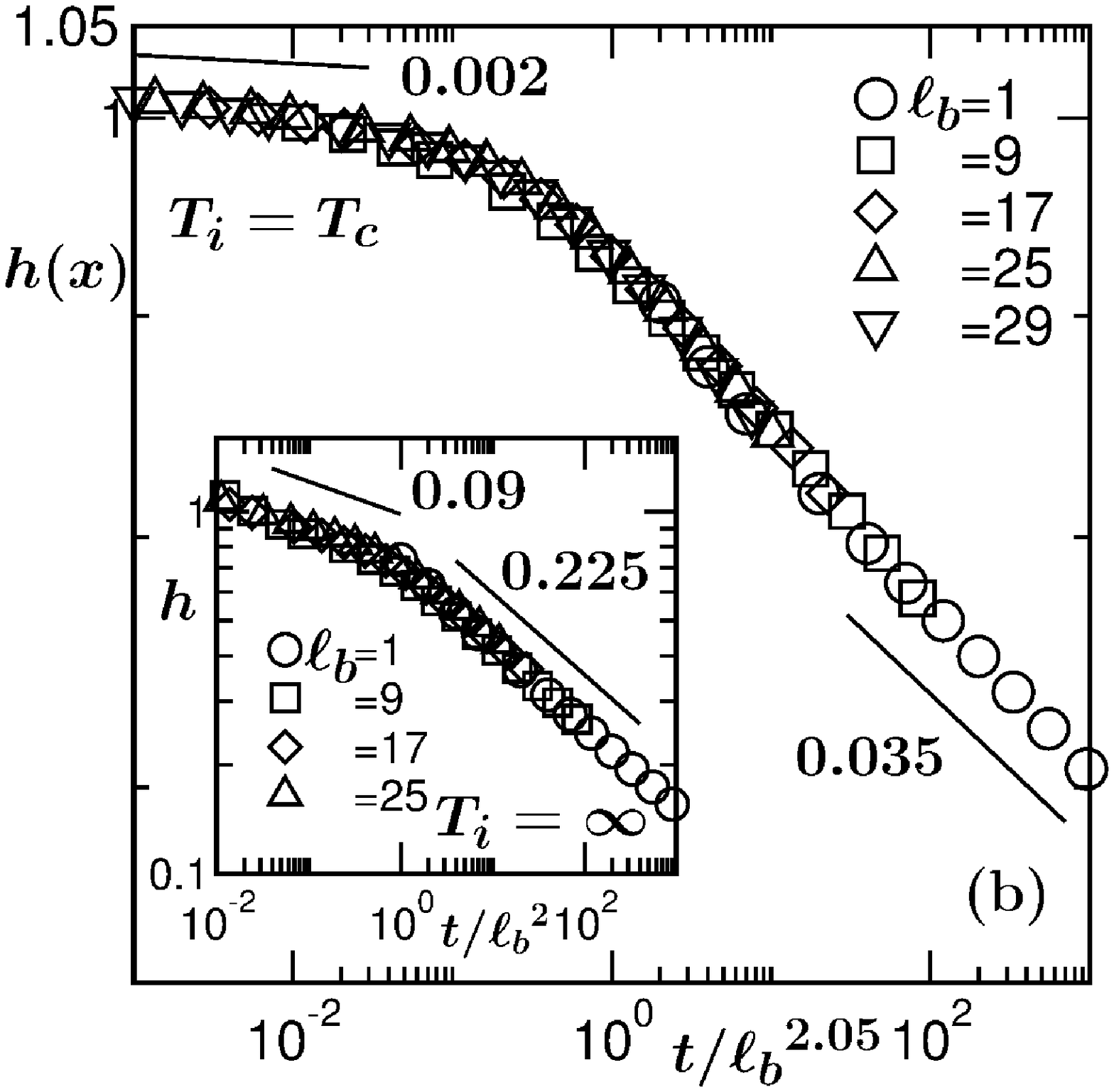}
\caption{\label{fig11}(a) Plots of block persistence probabilities, vs t,
from different values of $\ell_b$, for $T_i=T_c$, in $d=2$.
(b) Scaling plots of the persistence probabilities in (a). 
The scaling function $h(x)$ is plotted, 
on a log-log scale, vs $x=t/{\ell_b}^{1/\alpha}$. In the 
inset of (b) we present similar scaling plot for 
$T_i=\infty$. System sizes correspond to 
$L=2048$. Various power-law decays are shown by solid lines with the 
exponent values being mentioned next to appropriate lines.
}
\end{figure}

\par
In Fig. \ref{fig11}(a) we show $P_b$ vs $t$ plots from $d=2$, for a few different 
values of $\ell_b$ and $T_i=T_c$. It appears, as discussed, there exist two 
step decays and crossover to the faster (consistent with the local persistent 
decay) one is delayed with increasing $\ell_b$.
\par
In Fig. \ref{fig11}(b) we show a scaling 
exercise using the data of Fig. \ref{fig11}(a)  
where we have plotted $h(x)$ vs $t/{\ell_b}^{1/\alpha}$. For 
obtaining collapse of data, we have adjusted $\theta_0$ and $\alpha$. 
The value of $\alpha$ used here is $0.49$, 
that provides the best collapse. This number is certainly 
consistent with $1/2$, within numerical error. 
Early time behavior corresponds to global persistence with 
$\theta_{0}=0.002$ and the late time behavior is consistent with our previous 
estimation of $\theta_c \simeq 0.035$, for the site persistence probability. 
In the inset we have shown corresponding scaling results for $T_i= \infty$, for 
which $\theta_0$ and $\theta$ values (mentioned on the figure) 
are consistent with previous findings \cite{cueille_jpa}.
The value of $\alpha$ that provides the best collapse here is $0.5$. Note 
that in our earlier work such independence of $\alpha $ from $T_i$ was directly 
(from the analysis of $\ell$ vs t data) 
checked for this dimension. 
 
\section{\textrm{V} Conclusion}
We have presented results for coarsening dynamics 
in Ising model, with nonconserved order parameter, from space dimensions $d=2$ and $3$. 
The results include domain growth law and persistence, for quenches
with initial configurations of varying correlation length $\xi$. While presented 
results for persistence are mostly related 
to local order parameter \cite{satya_sire,satya_bray,%
derrida_hakim,derrida,stauffer}, for the global case \cite{cueille_jpa,cueille_epjb} we have 
obtained new exponent for quench from initial temperature $T_i=T_c$, in $d=2$. 
For local persistence, our results are summarized in the next paragraph.
\par
A central objective of this paper has been to identify the differences in 
the patterns formed by persistent spins when systems are quenched 
from $T_i=\infty$ and $T_i=T_c$, to the final temperature $T_f=0$. For both the
cases, corresponding fractal dimensionalities 
$d_f$, as well as the exponent $z$, 
related to the growth of the persistent pattern, have been obtained in various 
dimensions. A scaling law connecting $d_f$, $d$, $z$ and $\theta$, 
predicted by Manoj and Ray \cite{manoj_2}, has been 
observed to be valid, irrespective of the values of $d$ and $T_i$.
Combining various methods, we quote, for $T_i=\infty$, 
\begin{equation}\label{eq23}
\begin{split}
d_f=1.53\pm0.02,~ d=2, \\
d_f=2.65\pm0.03,~ d=3,
\end{split}
\end{equation}
and for $T_i=T_c$,
\begin{equation}\label{eq24}
\begin{split}
d_f=1.92\pm0.02,~d=2, \\
d_f=2.77\pm0.02,~d=3. 
\end{split}
\end{equation}
\par
On the standard domain growth problem, it  
is shown that the values of $\alpha$ in both dimensions 
for all initial temperatures are consistent with the theoretical 
expectation $\alpha=1/2$. This number describes the growth of the persistent 
pattern as well, validating Eq. (\ref{eq15}) and confirming that domain growth occurs 
essentially due to dynamics of spins along the domain boundaries. This resolves a controversy in $d=3$ for which 
some previous computer simulations reported $\alpha=1/3$. As mentioned 
in Ref. \cite{corberi}, this discrepancy must have been due to lack of data 
for extended period of time. Long simulations in our work, in addition 
to resolving this controversy, corrects the value of $\theta$ as well in
this dimension. 
\par
In future we will address similar issues for conserved order parameter dynamics, 
including aging phenomena. For both conserved and nonconserved dynamics, scaling 
properties and form of the two-point correlation function will be an important 
problem for the case of correlated initial configurations. 

\section*{Acknowledgement}\label{ack}
The authors thank Department of Science and Technology, Government of India, 
for financial support. SKD acknowledges hospitality and financial supports form
International Centre for Theoretical 
Physics, Italy. He is also thankful to the
Marie Curie Actions Plan of European Commision 
(FP7-PEOPLE-2013-IRSES grant No. 612707, DIONICOS).

~${*}$ das@jncasr.ac.in

\end{document}